\documentclass[english,11pt]{article}
\usepackage{epsf,amsmath,amssymb,graphicx,dcolumn}
\usepackage{scalefnt}
\usepackage{filemod}
\usepackage{subfigure}
\usepackage{pstricks}
\usepackage{cite,hyperref}
\usepackage{color}
\usepackage{rotating}
\newcommand{\muF}{\mu_\text{F}}
\newcommand{\muR}{\mu_\text{R}}

\newcommand{\mapproach}{$\qnt$-approach}
\newcommand{\qapproach}{$Q$-approach}
\newcommand{\mbottom}{m_b}
\newcommand{\mz}{M_Z}
\newcommand{\mhiggs}{M_H}
\newcommand{\siglo}{\hat{\sigma}^\text{\lo}_{\gghz}}
\newcommand{\zh}{Z\!H}
\newcommand{\gghz}{gg\to H\!Z}
\newcommand{\gensigma}{\Sigma}
\newcommand{\qnt}{M}
\newcommand{\genq}{W}
\newcommand{\genN}{N}
\newcommand{\gentau}{\tau}
\newcommand{\gentauhat}{\hat{\tau}}
\newcommand{\genc}{{\cal C}}
\newcommand{\tauq}{\tau_Q}

\newcommand{\taunt}{\tau_M}
\newcommand{\taunthat}{\hat{\tau}_M}
\newcommand{\api}{\frac{\as}{\pi}}
\newcommand{\dd}{\text{d}}

\newcommand{\as}{\alpha_{\mathrm{s}}}

\newcommand{\mtop}{m_t}

\newcommand{\tauhat}{\hat{\tau}}

\newcommand{\fig}[1]{Fig.\,\ref{#1}}
\newcommand{\eqn}[1]{Eq.\,(\ref{#1})}
\newcommand{\eqns}[1]{Eqs.\,(\ref{#1})}
\newcommand{\abbrev}{\scalefont{.9}}
\newcommand{\order}[1]{{\cal O}(#1)}
\newcommand{\lo}{{\abbrev LO}}
\newcommand{\nlo}{{\abbrev NLO}}
\newcommand{\nnlo}{{\abbrev NNLO}}
\newcommand{\nll}{{\abbrev NLL}}
\newcommand{\nnll}{{\abbrev NNLL}}
\newcommand{\nlonll}{{\abbrev NLO+NLL}}

\newcommand{\citere}[1]{Ref.\cite{#1}}
\newcommand{\citeres}[1]{Refs.\cite{#1}}
\newcommand{\sm}{{\abbrev SM}}
\newcommand{\qcd}{{\abbrev QCD}}
\newcommand{\lhc}{{\abbrev LHC}}
\newcommand{\pdf}{{\abbrev PDF}}

\title{\vspace*{-6em}
  \begin{flushright}
    {\small 
      October 2014\\[1em]
      MS-TP-14-17\\
      WUB/14-10\\
      LPN14-112\\
    }
  \end{flushright}
\vspace*{2em} {\bf 
Soft gluon resummation for gluon-induced Higgs Strahlung
}}
\author{ 
Robert V. Harlander$^a$,  Anna Kulesza$^{b}$, Vincent Theeuwes$^{b}$,
and Tom Zirke$^{a}$\\[2em]
 {\it $^a$Fachbereich C,
  Bergische Universit\"at Wuppertal,}\\[0em] {\it 42097 Wuppertal,
  Germany}\\[1em]
{\it $^b$Institute for Theoretical Physics, WWU M\"unster}\\[0em]
{\it D-48149 M\"unster, Germany}}
\date{}
\begin{document}
\maketitle


\vspace*{1cm}
\begin{abstract}
We study the effect of soft gluon emission on the total cross section
predictions for the $\gghz$ associated Higgs production process at the
\lhc. To this end, we perform resummation of threshold corrections at
the \nll{} accuracy in the absolute threshold production limit and in
the threshold limit for production of a $\zh{}$ system with a given
invariant mass. Analytical results and numerical predictions for various
possible \lhc{} collision energies are presented. The perturbative
stability of the results is verified by including universal \nnll{}
effects. We find that resummation significantly reduces the scale
uncertainty of the $\gghz$ contribution, which is the dominant source of
perturbative uncertainty to $\zh$ production.  We use our results to
evaluate updated numbers for the total inclusive cross section of
associated $pp \to Z\!H$ production at the \lhc. The reduced scale
uncertainty of the $\gghz$ component translates into a decrease of the
overall scale error by about a factor of two.
\end{abstract}
\vfill


\section{Introduction}

One of the main tasks for physics at the Run\,II of the \lhc{} will be
the precise determination of the properties of the newly found Higgs
boson at about 125\,GeV\cite{Aad:2012tfa,Chatrchyan:2012ufa}. Apart from
the enormous experimental activity behind the relevant measurements,
this requires a detailed theoretical understanding of the underlying
processes. Tremendous efforts have already been devoted in this
direction, the most central of which are collected in the three reports
of the \lhc{} Higgs Cross Section Working
Group\cite{Heinemeyer:2013tqa,Dittmaier:2012vm,Dittmaier:2011ti}.

\begin{figure}
  \begin{center}
    \begin{tabular}{cc}
      \includegraphics[viewport=64 690 180 775,
          width=.3\textwidth,page=1]{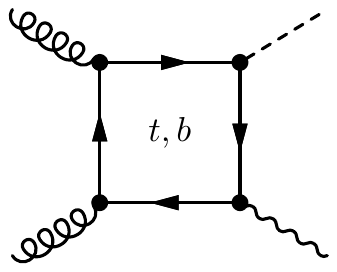}
      &
      \includegraphics[viewport=64 690 180 775,
          width=.3\textwidth,page=2]{figs/HZprodgraph.pdf}
    \end{tabular}
    \parbox{.9\textwidth}{%
      \caption[]{\label{fig:dias}Sample Feynman diagrams for the
        subprocess $\gghz$ to associated $\zh$ production.}}
  \end{center}
\end{figure}

Associated production of a Higgs boson with a weak gauge boson ({\it
  Higgs Strahlung} for short) may in fact turn out a key channel for
studying Higgs properties. The decay products of the final state gauge
boson allow for the separation of this process from the other production
channels, possibly with an additional restriction to the kinematical
regime of boosted Higgs bosons\cite{Butterworth:2008iy}.  An interesting
feature of Higgs Strahlung is that there are actually two variants of
it, namely $\zh{}$ and $W\!H$ production. While the next-to-leading
order (\nlo{}) \qcd{} corrections\cite{HVNLO,HVNLOrest} for these two
processes are identical, at next-to-\nlo{}
(\nnlo)\cite{NNLO,top,NNLOdiff} the $\zh$ process receives corrections
from gluon-initiated, quark-loop mediated contributions, which are
absent for $W\!H$ production, see \fig{fig:dias}. They are the main
subject of this paper and will be referred to as $\gghz$ in what
follows.\footnote{Note that at higher orders, other partonic initial
  states have to be taken into account as well in order to cancel
  initial state singularities.  However, as usual, we will still refer
  to them as $\gghz$ in this paper, unless indicated otherwise.}  In the
Standard Model (\sm{}), these corrections are completely dominated by
top- and bottom-quark loops and can amount to about 8\% in the total
cross section. In extended theories, this contribution can differ
significantly from the \sm{} prediction (see, e.g.,
\citeres{Kniehl:2011aa,Harlander:2013mla}).  One could therefore
consider the ratio $\sigma_{W\!H}/\sigma_{\zh}$ as a probe for new
physics, for example\cite{Harlander:2013mla}.


%
%


The $\gghz$ contribution induces a significant uncertainty to the $\zh$
cross section (and therefore also to the $W\!H/\zh{}$ ratio), since
their renormalization and factorization scale dependence is not
compensated by any lower order terms.

The \nlo{} \qcd{} corrections to this contribution
(i.e.\,$\order{\as^3}$) have been evaluated in
\citere{Altenkamp:2012sx}, albeit in the heavy-top limit, i.e., by
letting $m_t\to \infty$ and $m_b\to 0$. They increase the
gluon-initiated contribution by roughly 100\%; the \nlo{} result still
suffers from a scale uncertainty of 20-30\%, which corresponds to about
2-3\% relative to the {\it total} $\zh$ cross section.

The gluon-induced $\zh$ process bears a striking similarity in the
production mechanism to the well-known gluon fusion process $gg\to H$
(see, e.g.\,\citere{Djouadi:2005gi}), also manifested in the numerical
behaviour of the two processes. The substantial size of the \nlo{}
corrections ($K$-factors of two) can be attributed to the soft gluon
emission off the initial state, leading to logarithmic contributions
which get large in the threshold limit for the production. By performing
resummation of these corrections, one can not only take into account
these logarithmic contributions to all orders in perturbation theory,
but also reduce the theoretical uncertainty due to scale variation. It
is the purpose of this paper to perform this resummation at the
next-to-leading logarithmic (\nll{}) accuracy and match the resummed
result to the \nlo{} predictions, obtaining in this way the \nlonll{}
result. We will also discuss the effect of the next-to-\nll{} (\nnll{})
corrections on the theoretical predictions.
    
Threshold resummation techniques have been used extensively in the
recent years to improve the knowledge of Higgs production cross
sections. The most prominent example is the inclusive Higgs production
via gluon fusion, $gg \to H$\cite{Catani:2003zt,Kulesza:2003wn,%
  Moch:2005ky,Laenen:2005uz,
  Ravindran:2005vv,Idilbi:2005ni,Ahrens:2008qu,Ball:2013bra,Bonvini:2014joa}.
Resummation of threshold logarithms has also been used to improve the
predictions for the $gg\to H$ process when the Higgs is produced at high
transverse momentum and the accompanying jet remains
unobserved\cite{deFlorian:2005rr, Huang:2014mca, Becher:2014tsa}. For
the process of interest here, i.e. the associated Higgs production with
a weak gauge boson, threshold resummation has been performed so far only
for the $q \bar q$ initiated partonic processes\cite{Dawson:2012gs}. For
an extensive discussion of the subject we refer the reader
to\,\citeres{Heinemeyer:2013tqa,Dittmaier:2012vm,Dittmaier:2011ti} and
references therein.

In contrast to the process $gg\to H$ where the threshold is uniquely
defined at the partonic center-of-mass energy $\hat s=\mhiggs^2$, one
may argue along two different lines in $\zh$ production. On the one
hand, the {\it absolute} threshold for this process is given by the
minimum of the invariant mass of the $\zh$ system, $\hat
s_\text{min}=(\mhiggs+\mz)^2$. Indeed, the total cross section receives
a large portion from the kinematical region around this threshold,
partly due to the usual logarithmic enhancement from soft and collinear
gluon radiation, but also due to the gluon luminosity which drops quite
steeply towards larger values of the partonic center-of-mass energy. On
the other hand, one may consider the production of the $\zh$ system at a
definite value of the squared invariant mass $Q^2\equiv (p_H+p_Z)^2$,
and resum the threshold logarithms with respect to that scale. From a
formal point of view, this latter approach is in closer relation to the
one for the Drell-Yan process considered at a fixed invariant mass of
the lepton pair, or the $gg\to H$ process at a given Higgs mass.

We are going to present the formalism and the results for both of these
approaches and discuss the differences between them. The behavior of the
final results will allow us to define a preferred approach, in the sense
that it allows for a more appropriate scale choice and therefore appears
to be perturbatively better behaved. Both approaches lead to compatible
results, however, which provides us with an important consistency check.

The remainder of the paper is structured as follows: In the next section
we discuss the threshold resummation formalism, whereas in
Section~\ref{sec:lo} we present process-specific results needed to
perform resummation at \nll{}. Section~\ref{sec:nllresults} is dedicated
to the discussion of the numerical \nlonll{} predictions at the \lhc. It
also briefly addresses the universal part of the \nnll{}
effects. Finally, in Section~\ref{sec:update} we update the total
inclusive cross section predictions of associated $\zh$ production at
the LHC; we conclude in Section~\ref{sec:conclusions}.


\section{Formalism}\label{sec:formalism}

Let us consider the hadronic cross section $\gensigma$ for the
production of a color-neutral system which involves a scale $\genq$,
defined as a characteristic scale for the production threshold,
cf.~\citeres{Sterman:1986aj,Catani:1989ne,Czakon:2009zw}. In the
following we will consider a $2 \to 2$ process in a pair-invariant
kinematics, whenever appropriate. With the help of the hadronic and
partonic threshold variables
\begin{equation}
\begin{split}
\gentau \equiv \genq^2/s\,,\qquad
\gentauhat \equiv \genq^2/\hat s\,,\qquad
\end{split}
\end{equation}
where $s$ and $\hat s$ are the hadronic and the partonic center-of-mass
energy, respectively, this cross section can be written as
\begin{equation}
\begin{split}
\gensigma(\genq^2,\gentau) \!= \!\sum_{i,j} \!\!\int\!\! \dd x_1 \dd x_2 \, d
\gentauhat \; \delta\left(\gentauhat - \frac{\gentau}{x_1 x_2}\right)
f_{i}(x_1,\mu ) f_{j}(x_2,\mu )
\,\hat\gensigma_{ij}(\genq^2,\gentauhat,\mu^2),
\label{eq:genxsection}
\end{split}
\end{equation}
where we have set the renormalization and factorization scales equal to
$\mu$ in order to simplify the discussion. The dependence on all
relevant mass parameters, other than that related to the threshold
scale, is implicit in Eq.~\ref{eq:genxsection} and throughout the rest
of this paper.

At higher orders in perturbation theory, the partonic cross section
$\hat\gensigma_{ij}$ contains logarithmic terms originating from soft
gluon emission that diverge in the threshold limit, i.e.\ for
$\tauhat\to 1$.
Such terms clearly spoil the perturbative behavior of a na\"ive
expansion in $\as$ in the threshold region. However, these logarithmic
terms can be treated systematically with the help of threshold
resummation which is most conveniently formulated in Mellin space, where
the hadronic cross section reads
\begin{equation}
\begin{split}
\label{eq:Nspacexsec}
\tilde\gensigma(\genq^2,\genN)&\equiv \int_0^1 \dd \gentau
\gentau^{\genN-1} \gensigma_{ij}(\genq^2,\gentau) \\ &=
 \sum_{i,j}
\tilde f_{i} (\genN+1,\mu^2)\,\tilde f_{j} (\genN+1, \mu^2) \,
\tilde{\hat\gensigma}_{ij}(\genq^2,\genN,\mu^2)\,.
\end{split}
\end{equation}
The Mellin transform of a function $h(\ldots, x,\ldots)$ w.r.t.\ the
variable $x$ is defined as usual as
\begin{equation}
\begin{split}
\tilde{h} (\ldots,N,\ldots)
\,\equiv\, \int_0^1 dx \, x^{N-1} \, h(\ldots,x,\ldots)\,.
\label{eq:mellin}
\end{split}
\end{equation}
The moments of the hadronic (partonic) cross section are taken
w.r.t.\ the variable $\tau$ ($\tauhat$), those of the parton
distributions $f_{i}(x,\mu^2)$ are taken w.r.t.\ the partonic momentum
fraction $x$.  Mellin transformation turns the terms that are
logarithmic in $1-\tau$ (respectively $1-\tauhat$) into logarithms of
the Mellin variable $N$; it is the latter which are then resummed to all
orders in $\as$.

As the subject of this paper is the gluon-initiated component of
associated $\zh$ production, $\sigma(\gghz)$, it will be convenient to
specialize the presentation to gluon-initiated processes in the
following. The resummed partonic cross section then has the
form:
\begin{equation}
\begin{split}
\tilde{\hat\gensigma}^\text{(res)}(\genq^2,\genN,\mu^2)\,&=
\tilde{\hat\gensigma}^\text{\lo}(\genq^2,\genN,\mu^2) \, {\cal
  C}_{\genq}(\genq^2,\mu^2)\\&\qquad\times\Delta_g(\genq^2,\genN+1,\mu^2)
\Delta_{g}(\genq^2,\genN+1,\mu^2)\,.
\label{eq:resumm}
\end{split}
\end{equation}
The \lo{} partonic cross section, $\hat\gensigma^\text{\lo}$, is
obtained from Feynman diagrams with closed top- and bottom-quark loops,
see \fig{fig:dias}.\footnote{Note that the bottom-loop contributions
  could lead to a breaking of factorizaton at $1-\tauhat\sim
  m_b^2/W^2$ which would be in conflict with
  \eqn{eq:resumm}. However, in \citere{Altenkamp:2012sx} it was shown
  that these diagrams are absent for $m_b=0$ in Landau gauge. The
  genuine bottom-loop effects are thus suppressed by the bottom Yukawa
  coupling or by powers of $m_b/\qnt$. Using {\tt
    vh@nnlo}\cite{Brein:2012ne}, we checked that their numerical effect
  is at the percent level, so that this issue can safely be ignored for
  the scope of this paper. We note in passing that also for the gluon
  fusion process, $gg\to H$, soft gluon resummation for the bottom-loop
  induced terms has been ignored so far. For a related discussion in the
  context of small-$p_T$ resummation, see
  \citeres{Mantler:2012bj,Grazzini:2013mca,Banfi:2013eda,Harlander:2014uea}.}.
The Sudakov factor $\Delta_g$ describes the soft emission off an
incoming gluon $g$; in the $\overline{\mathrm{MS}}$ factorization scheme
it is given by, up to the \nnll{} level,
\begin{equation}
\begin{split}
& \log \Delta_{g}(\genq^2,N,\mu^2) =\\&\qquad = \int_0^1 dz
\,\frac{z^{N-1}-1}{1-z} \int_{\mu^2}^{ \genq^2(1-z)^2} \frac{dq^2}{q^2}
A_g(\as(q^2))+D_g(\as((1-z)^2\genq^2))\,,
\label{eq:sudakov}
\end{split}
\end{equation}
where the resummation coefficients have the following perturbative
expansions:
\begin{equation}
\begin{split}
A_g(\as) &= \sum_{n\geq 1} \left(\api\right)^n A_g^{(n)}\,,\qquad
D_g(\as) = \sum_{n\geq 2} \left(\api\right)^n D_g^{(n)}\,.
\end{split}
\end{equation}
Through \nll{}, only $A_g$ contributes, with the leading logarithmic
({\abbrev LL}) and \nll{} coefficients given by
\cite{Kodaira:1981nh,Catani:1990rr} ($C_A=3$, $T_R=1/2$)
\begin{equation}
\begin{split}
A_g^{(1)} &= C_A\,,\qquad A_g^{(2)} =
C_A^2\left(\frac{67}{36}-\frac{\pi^2}{12}\right) -\frac{5}{9}C_AT_R\,.
\label{eq:ag}
\end{split}
\end{equation}
At \nnll{} accuracy one requires the knowledge of $D_g^{(2)}$ and
$A_g^{(n)}$ through $n=3$.  Explicit results for these coefficients can
be found in \citere{Moch:2005ba,Vogt:2000ci,Catani:2001ic}. In numerical
applications, we use expansions of \eqn{eq:sudakov} up to \nll{},
respectively \nnll{} terms in $N$, see Eqs.\,(24) and (27) of
\citere{Catani:2003zt}.

The final ingredient in \eqn{eq:resumm} is the process dependent
perturbative function $\genc_\genq$, which contains the ``hard''
contributions, i.e.\ terms that are constant in the large $N$ limit. At
\nll{}, they originate from \nlo{} virtual and the non-logarithmic soft
corrections. Correspondingly, $\genc_\genq$ requires the knowledge of
the threshold limit of the \nlo{} cross section.  It is well known that
the latter can be cast into the form
\begin{equation}
\hat\gensigma^\text{\nlo}= \int_3 \left[ \dd
  \hat\gensigma^{\rm R}|_{\epsilon=0} - \dd \hat\gensigma^{\rm
    A}|_{\epsilon=0} \right]+\int_2 \left[\dd \hat\gensigma^{\rm V} +
  \int_{1} \dd \hat\gensigma^{\rm A} \right]_{\epsilon=0} +
\hat\gensigma^{\rm C}
\end{equation}
using the standard notation of $\dd\hat\gensigma^{\rm R}$ for the real
part of the corrections, $\dd\hat\gensigma^{\rm V}$ for the virtual
part, $\dd\hat\gensigma^{\rm A}$ for the auxiliary singular dipole
terms\cite{Catani:1996vz} and $\hat\gensigma^{\rm C}$ for the collinear
counter terms. By construction of the dipole terms, the first integral
on the r.h.s.\ vanishes faster in the soft limit than the remaining
terms and therefore can be ignored. We have also confirmed this
numerically using the specific matrix elements and subtraction terms.
The remaining terms can be written as
\begin{equation}
\begin{split}
\hat\gensigma^\text{\nlo}(\tauhat) = \hat\gensigma^V_\text{reg}(\tauhat)
+ 2\int_{\gentauhat}^1\dd x\;[\textbf{\textit{P}}^{gg}
  +\textbf{\textit{K}}^{gg}](x)
\;\hat\gensigma^\text{\lo}\left(\gentauhat/x\right) +\ldots\,,
\label{eq:signlosoft}
\end{split}
\end{equation}
where the ellipse denotes terms that are sub-leading in the soft limit,
and\cite{Catani:1996vz}
\begin{equation}
\begin{split}
&\left[\textbf{\textit{P}}^{gg}+\textbf{\textit{K}}^{gg}\right](x)
  =\\&\qquad= \frac{\as}{2\pi}\left(
  2C_A\left\{2\left(\frac{\log(1-x)}{1-x}\right)_{+}
  -\left(\frac{1}{1-x}\right)_{+}\log\left(\frac{\mu^2}{\hat{s}}\right)\right.\right.\\ &\quad\qquad-\left[\frac{1-x}{x}-1+x(1-x)\right]\log\left(\frac{\mu^2}{(1-x)^2\hat{s}}
  \right)-\left.\delta(1-x)\frac{\pi^2}{6}\right\}\\ &\qquad\quad\left. -\delta(1-x)\left\{\left(\frac{11}{6}C_A-\frac{2}{3}T_R\;n_l\right)
  \log\left(\frac{\mu^2}{\hat{s}}\right)+\left(\frac{50}{9}-\frac{2\pi^2}{3}
  \right)C_A-\frac{16}{9}T_R\;n_l\right\}\right)
\end{split}
\end{equation}
denotes process independent terms, while $\hat\gensigma^V_\text{reg}$
are the {\abbrev IR}-subtracted virtual corrections to the total \nlo{}
cross section according to the method of dipole
subtraction\cite{Catani:1996vz}. For loop-induced processes they
typically require a genuine two-loop calculation.

Note that in Mellin space, the second term on the r.h.s.\ of
\eqn{eq:signlosoft} factorizes into the Mellin transforms of
$\hat\gensigma^\text{\lo}(x)$ and
$x\left[\textbf{\textit{P}}^{gg}+\textbf{\textit{K}}^{gg}\right](x)$.
The latter is easily evaluated using well-known relations for Mellin
transforms (see, e.g., \citere{Blumlein:1998if}), and in the large $N$
limit consists of terms that are logarithmic or constant in $N$.  For
the terms constant in $N$ (as $N\to \infty$) which is all that is needed
here, we find
\begin{equation}
\begin{split}
U(\genq^2) &\equiv 2\int\dd x\,x^{N-1}
x\left[\textbf{\textit{P}}^{gg}+\textbf{\textit{K}}^{gg}\right](x)\bigg|_\text{$N$-const}
=\\&= \frac{\as}{\pi}\bigg\{
  \left[\frac{2}{3}T_R\;n_l-\left(\frac{11}{6}-2\gamma_E\right)C_A\right]
  \log\left(\frac{\mu^2}{\genq^2}\right)\\&\qquad
  -\left(\frac{50}{9}-\frac{2\pi^2}{3} - 2\gamma_E^2
  \right)C_A+\frac{16}{9}T_R\;n_l\bigg\}\,,
\label{eq:universal}
\end{split}
\end{equation}
where, as usual, $\gamma_E = 0.5772157\ldots$ denotes Euler's
constant. \eqn{eq:universal} is in agreement with the expression for the
$\tilde{I}^{(1)}_g$ operator introduced in \citere{Catani:2013tia} after
accounting for terms originating from the Mellin transform and adding
universal contributions containing $1/\epsilon$ poles in the dipole
subtraction method.

The validity of the threshold resummed cross section
$\hat\gensigma^\text{(res)}$ is restricted to the soft region
$\gentauhat\approx 1$. In order to arrive at a result which is valid for
general $\gentauhat$, one should combine the resummed and the
fixed-order expression without double counting terms that are contained
in both. This is achieved by subtracting the perturbative expansion of
the resummed cross section.  The resummation-improved hadronic cross
section therefore reads:
\begin{equation}
\begin{split}
\gensigma^\text{(f.o.+l.a.)}(\genq^2,\gentau) & = \gensigma^\text{\abbrev
  (f.o.)}(\genq^2,\gentau) + \sum_{ij}\int_\mathrm{\abbrev CT} \dd\genN \;
\gentau^{-\genN}\; \tilde f_{i} (\genN+1, \mu^2) \; \tilde f_{j}
(\genN+1,\mu^2) \\ & \times\, \left[
  \tilde{\hat\gensigma}_{ij}^\text{(l.a.)} (\genN, \mu^2) -
  \tilde{\hat\gensigma}_{ij}^\text{(l.a.)} (\genN, \mu^2)
  \bigg|_\text{\abbrev (f.o.)}\, \right]\,,
\label{eq:match}
\end{split}
\end{equation}
where ``(f.o.)'' denotes the fixed order result, while ``(l.a.)'' marks
the resummed expression. The inverse Mellin transform can be evaluated
numerically using a contour {\abbrev CT} in the complex-$N$ space
according to the ``Minimal Prescription'' method developed in
\citere{Catani:1996yz}.

As outlined in the introduction of this paper, we will pursue two
alternative approaches for the resummation of threshold logarithms in
the $\gghz$ process which essentially differ by the specific choice for
$\genq$.

In the first approach, dubbed {\it \qapproach{}} in the following, we
will consider threshold resummation for the production of the $\zh{}$
system at a {\it fixed invariant mass} $Q$, i.e., we are going to set
\begin{equation}
\begin{split}
\genq&=Q\equiv\sqrt{(p_H+p_Z)^2}\,,\qquad \gentau=\tauq\equiv Q^2/s\,,\\
\gensigma(\genq^2,\gentau) &= \frac{\dd\sigma_{\gghz}}{\dd
  Q^2}\,,
\end{split}
\end{equation}
and analoguously for the partonic quantities.  $\sigma_{\gghz}$ is the
cross section for the process $\gghz$, and $p_Z$ and $p_H$ are the final
state four-momenta. The logarithmic terms which are treated by the
resummation formalism are of the form
\begin{equation}
\begin{split}
\as^{2+n}\left(\frac{\log^m(1-\tauhat_Q)}{1-\tauhat_Q}\right)_+\,,\quad
m\leq 2n-1\,,
\end{split}
\end{equation}
in this case. The $+$-distribution is defined in the usual way as
\begin{equation}
\begin{split}
\int_0^1\dd\tauhat_Q\, g(\tauhat_Q) [f(\tauhat_Q)]_+ 
\equiv \int_0^1\dd\tauhat_Q
\left[g(\tauhat_Q)-g(1)\right] f(\tauhat_Q)\,.
\label{eq:plusdist}
\end{split}
\end{equation}
The total cross section can be
obtained upon subsequent integration over $Q^2$.

In the second approach, we will consider the ($Q^2$-integrated) total
inclusive cross section for the process $\gghz$, and resum the
logarithms relative to the {\it absolute threshold} for the production
of the $\zh{}$ pair, $\qnt^2 = (\mhiggs+\mz)^2$. Correspondingly, we
will refer to this procedure as {\it \mapproach{}} and set
\begin{equation}
\begin{split}
\genq&=\qnt\equiv \mhiggs+\mz\,,\qquad 
\gentau=\taunt\equiv \qnt^2/s\,,\\
\gensigma(\genq^2,\gentau) &= \sigma_{\gghz}\,.
\end{split}
\end{equation}
In this case, the divergences at threshold occur in the form
\begin{equation}
\begin{split}
\as^{2+n}\log^m(1-\tauhat_M)\,,\qquad m\leq 2n\,.
\end{split}
\end{equation}
The difference between the two approaches has been studied in the
literature in the context of $t \bar t$ production\cite{Czakon:2009zw,
  Ahrens:2010zv, Beneke:2011mq}. In particular, in
\citere{Czakon:2009zw} it has been shown that the total cross section
obtained by integrating the \qapproach{} result over the invariant mass
contains the same leading logarithmic terms as those in the
\mapproach{}, but differs by power-suppressed terms. In what follows, we
pursue the calculation in both approaches and treat them in a
complimentary way. In practice, however, since the \qapproach{} allows
for a $Q$-dependent choice of the renormalization and factorization
scale, it appears to be better suited to describe physics of the process
at hand.


\section{Calculation of the resummed cross sections}\label{sec:lo}

Let us now discuss the various specific ingredients that enter the
expressions for the threshold resummed cross sections at \nll{} accuracy
in the two approaches.


\subsection{The invariant mass threshold limit: \qapproach{}}

At \lo{}, the partonic invariant mass distribution is proportional to a
$\delta$ function,
\begin{equation}
\begin{split}
\label{eq:loq}
\frac{\dd\siglo}{\dd Q^2} =
\siglo(\hat s,\mhiggs,\mz)\,\delta(\hat s-Q^2)\,,
\end{split}
\end{equation}
where $\siglo$ is the total inclusive cross
section for $\zh{}$ production at
\lo{}\cite{gghzlo,Brein:2003wg}.
The Mellin transform is therefore constant in $N$:
\begin{equation}
\begin{split}
\frac{\dd\tilde{\hat{\sigma}}^\text{\lo}_{\gghz}}{\dd Q^2} =
\frac{\siglo(\hat s=Q^2,\mhiggs,\mz)}{Q^2}\,.
\end{split}
\end{equation}
Note that due to the form of the \lo{} cross section in \eqn{eq:loq},
the convolution in \eqn{eq:signlosoft} turns into a simple product
proportional to the \lo{} cross section ${\siglo(Q^2,\mhiggs,\mz)}$ and
the term
$\left[\textbf{\textit{P}}^{gg}+\textbf{\textit{K}}^{gg}\right](x)$
taken at $x=\hat \tau_Q$.  The exact partonic \lo{} cross section
$\siglo$ is implemented in {\tt
  vh@nnlo}\cite{Brein:2012ne,Harlander:2013mla} in terms of
Passarino-Veltman functions which are evaluated with the help of {\tt
  LoopTools}\cite{Hahn:2000jm}. We use the corresponding subroutines of
{\tt vh@nnlo} for our numerical analysis.

Obviously, the kinematics of the virtual corrections is the same as for
the \lo{} cross section, and we can write
\begin{equation}
\begin{split}
\hat{\tilde\gensigma}^V_\text{reg}(Q^2,N) =
\frac{\hat\sigma_\text{virt}}{Q^2}\,,\label{eq:sigvirt}
\end{split}
\end{equation}
where the virtual cross section at \nlo{} has been evaluated in the
limit of an infinitely heavy top quark\cite{Altenkamp:2012sx}:
\begin{equation}
\begin{split}
\hat\sigma_\text{virt} &= K(\hat s)\cdot\siglo +
\hat\sigma_\text{(virt,red)}\,,\\[1em]
\text{with}\quad K(\hat s)&=1 + \frac{\as(\mu)}{\pi}\left(
  \frac{164}{9} + \frac{23}{6}\ln\frac{\mu^2}{\hat
    s}\right)\,.
\label{eq:virt}
\end{split}
\end{equation}
$\siglo$ is again the total inclusive partonic \lo{} cross section,
while the ``reducible'' contribution $\hat\sigma_\text{(virt,red)}$ is
due to Feynman diagrams that involve two quark triangles (see
\citere{Altenkamp:2012sx} for details). It was calculated in the
heavy-top limit in \citere{Altenkamp:2012sx}; we use the resulting
compact analytic expression in our numerical analysis. In order to
approximate the $m_t$ dependence, we rescale it by
$\siglo/\hat{\sigma}^\text{LO}_\infty$, where the \lo{} cross section in
the heavy-top limit is given by\cite{Altenkamp:2012sx}
\begin{equation}
\begin{split}
\lim_{\mtop\to\infty}\siglo(\hat s,\mhiggs,\mz)
\equiv
\hat{\sigma}^\text{\lo}_{\infty}(\hat s,\mhiggs,\mz) = \kappa\cdot
\frac{\lambda^{3/2}(\hat{s},\mhiggs^2,\mz^2)}{\mz^4\hat{s}^2}\,,
\label{eq:gghzloinf}
\end{split}
\end{equation}
with
\begin{equation}
\begin{split}
\lambda(x,y,z) = x^2+y^2+x^2-2xy-2yz-2zx\,.
\end{split}
\end{equation}
The factor
\begin{equation}
\begin{split}
\kappa &=
\frac{1}{\pi}\left(\frac{\as\alpha}
{64\,c^2_\text{W}s^2_\text{W}}\right)^2
\label{eq:norm}
\end{split}
\end{equation}
collects numerical constants that are not directly relevant for the
discussion in this section.

Up to $\cal{O}(\as)$, the hard coefficient in the \qapproach{} then
reads
\begin{equation}
\begin{split}
{\cal C}_{Q} &=
K(Q^2) + U(Q^2) + \hat\sigma_\text{(virt,red)}/
\hat{\sigma}^\text{LO}_\infty\bigg|_{\hat s=Q^2}\,,
\label{eq:calcq}
\end{split}
\end{equation}
with the universal contribution $U$ defined in
\eqn{eq:universal}.\footnote{\label{fn:c1gga}We note in passing that,
  apart from the reducible contribution $\sigma_\text{(virt,red)}$, the
  \nlo{} hard coefficient coincides with the one for pseudo-scalar Higgs
  production in gluon fusion, see \citere{deFlorian:2007sr}. This is
  expected since it was observed that the corresponding amplitudes for
  the two processes are the same in the
  heavy-top limit\cite{Altenkamp:2012sx}.}


\subsection{The absolute threshold limit: \mapproach{}}

The \mapproach{} requires the Mellin transform of the \lo{} total inclusive
cross section $\siglo$ w.r.t.\ the variable $\taunthat = \qnt^2/\hat s$,
which is non-trivial to obtain due to the complicated dependence on
$\qnt = \mhiggs+\mz$. Since the \mapproach{} is concerned with the
logarithms close to the $Z\!H$ threshold where $\hat s< 4\mtop^2$, it is
justified to apply the heavy-top limit though, given in
\eqn{eq:gghzloinf}.

Note, however, that $\hat\sigma^\text{\lo}_\infty$ diverges as $\hat s\to
\infty$. The Mellin integral w.r.t.\ $\taunthat=\qnt^2/\hat s$ therefore
does not exist. This is not surprising, since the integration region of
the Mellin transform extends to values of $\hat s=\qnt^2/\taunthat$
where the heavy-top limit does not hold.  In its validity range,
however, $\hat{\sigma}^\text{LO}_\infty$ can safely be approximated by its
threshold expansion:
\begin{equation}
\begin{split}
\hat{\sigma}^\text{\lo}_{\infty}(\hat s,\mhiggs,\mz)&\stackrel{\hat s\to
  \qnt^2}{\to} \hat{\sigma}^\text{\lo}_{\infty,\mathrm {thr.}}(\hat
s,\mhiggs,\mz) = \frac{8\kappa}{r_M\mz^2}
(r_M-1)^{3/2}\,(1-\taunthat)^{3/2}\,,\\
\mbox{where}\quad& r_M \equiv 1+\mhiggs/\mz = M/\mz\,,
\label{eq:LOthresh}
\end{split}
\end{equation}
and higher terms in $(1-\taunthat)$ have been dropped. Note that the
\lo{} cross section tends to zero quite fast at threshold, with one
power of $(1-\taunthat)^{1/2}$ due to the phase space, and two
additional powers arising from the squared amplitude.  The reason for
this behaviour is that at threshold, both the Higgs and the $Z$ are at
rest and therefore the $Z\!H$ system has orbital angular momentum
zero. The total angular momentum of the final state is thus given by the
spin of the $Z$ boson, $j=1$. Such a configuration of final and initial
states is forbidden by the Landau-Yang theorem, and therefore the
amplitude must vanish at threshold.

The expression in \eqn{eq:LOthresh} can easily be transformed into
Mellin space using
\begin{equation}
\begin{split}
\int_0^1\dd \taunthat \,\taunthat^{N-1} (1-\taunthat)^{3/2} = 
\frac{3\sqrt{\pi}}{4}\frac{\Gamma(N)}{\Gamma(N+5/2)}\,,
\label{eq:lomellin}
\end{split}
\end{equation}
with Euler's $\Gamma$-function obeying $z\Gamma(z)=\Gamma(z+1)$ and
$\Gamma(1)=1$.

For the determination of the hard matching coefficient ${\cal C}_\qnt$,
we also require the threshold limit (defined in Mellin space) of the
\nlo{} virtual corrections. In the \mapproach{}, the latter are
directly given by \eqn{eq:virt}. Having already dealt with $\siglo$
(the term $\ln(\mu^2/\hat s)$ in $K(\hat s)$ of \eqn{eq:virt} can
simply be replaced by $\ln(\mu^2/\qnt^2)$ in the threshold limit), what
remains is to evaluate the Mellin transform of the reducible
terms. Following the same line of argument as for $\siglo$, we take the
threshold limit of the heavy-top limit expression given in
\citere{Altenkamp:2012sx}:
\begin{equation}
\begin{split}
\hat\sigma_\text{(virt,red)} &\stackrel{\taunthat\to 1}{\to}
 \hat\sigma_\text{(virt,red),thr.} =\\&=
-\frac{2}{9}\api\,\hat{\sigma}^\text{\lo}_{\infty,\text{thr}}(\hat
s,M_H,M_Z)
\left[1+r_M+\frac{2+r_M}{2r_M}\ln(r_M-1)\right]\,.
\label{eq:virtred}
\end{split}
\end{equation}
Note that the threshold behavior of the virtual
corrections is the same as for the \lo{} cross section, given by
$\hat{\sigma}^\text{\lo}_{\infty,\text{thr}}$ defined in
\eqn{eq:LOthresh}. Correspondingly, the Mellin transform can be obtained
again using \eqn{eq:lomellin}.

The hard coefficient in the \mapproach{} therefore follows as
\begin{equation}
\begin{split}
{\cal C}_{\qnt} &= K(\qnt^2) + U(\qnt^2) +
\hat\sigma_\text{(virt,red),thr.}/\hat\sigma^\text{\lo}_{\infty,\text{thr.}}\,,
\label{eq:cm}
\end{split}
\end{equation}
with the functions $K$ and $U$ defined in \eqns{eq:virt} and
(\ref{eq:universal}), respectively. Note that the $\taunthat$-dependence
cancels in the last term on the r.h.s.\ of \eqn{eq:cm}.  The final
resummed hadronic cross section in the \mapproach{} is rescaled by
$\sigma^\text{\lo}_{\gghz}/\sigma^\text{\lo}_{\infty,\,\text{thr}}$, in
a similar manner to the \nlo{} calculation~\cite{Altenkamp:2012sx}.



\section{Results for the resummed cross section}\label{sec:nllresults}

In this section, we are going to study the numerical impact of the
resummed terms on the inclusive total cross section at \nlonll, i.e., in
\eqn{eq:match}, we set ``(f.o.)''=\nlo, and ``(l.a.)''=\nll.  Particular
focus will be put on the theoretical uncertainty which is estimated in
the usual way by varying the renormalization and the factorization
scales around a default value $\mu_0$. Furthermore, \nnll{} effects will
be discussed at the end of this section.

At fixed order, $\mu_0$ is typically chosen as the invariant mass of the
$\zh$ system (see, e.g., \citere{Brein:2003wg}), since this is the
characteristic scale of the partonic process. In the \qapproach{}, we are
going to adopt this choice, which allows us to directly compare to
previous predictions of the total cross section. For the \mapproach{},
however, the invariant mass is integrated over before the resummation is
performed; this scale is therefore unaccessible as central choice for
$\muF$ or $\muR$. In fact, since this approach emphasizes the {\it
  absolute} threshold $\qnt=\mz+\mhiggs$, it is this scale that should
be considered as default in the \mapproach.

The other numerical parameters entering the total cross section are
chosen as in \citere{Heinemeyer:2013tqa}. In particular, we use $\mtop =
172.5$\,GeV for the on-shell top mass, and $\mbottom=4.75$\,GeV for the
on-shell bottom mass.  Concerning the \pdf{} set, it is not obvious {\it
  a priori} which order is appropriate for the $\gghz$ process. From a
renormalization group and {\abbrev DGLAP} point of view, the $\gghz$
subprocess adds incoherently to the Drell-Yan like terms, so that its
$\as^2$ contribution could be considered as \lo{}. The numerical results
of \citere{Altenkamp:2012sx} were produced by adopting this point of
view. On the other hand, the lowest order contribution to $\gghz$ is
$\order{\as^2}$ which is the same as for the \nnlo{} prediction of the
total inclusive Higgs Strahlung process; one could therefore consider it
as \nnlo{}. This is the viewpoint taken in \citere{Heinemeyer:2013tqa},
and it will also be our default choice.  We therefore evaluate all
hadronic cross sections using \nnlo{} \pdf{} sets; specifically, we use
{\abbrev MSTW}2008{\abbrev NNLO}.\footnote{Note that from this
  perspective, the \nlo{} and \nlonll{} prediction for $\gghz$ should be
  evaluated using even higher order \pdf{} sets, which are, however,
  unavailable to date.}


%
\begin{figure}
  \begin{center}
    \begin{tabular}{cc}
      \includegraphics[height=.35\textheight]{%
        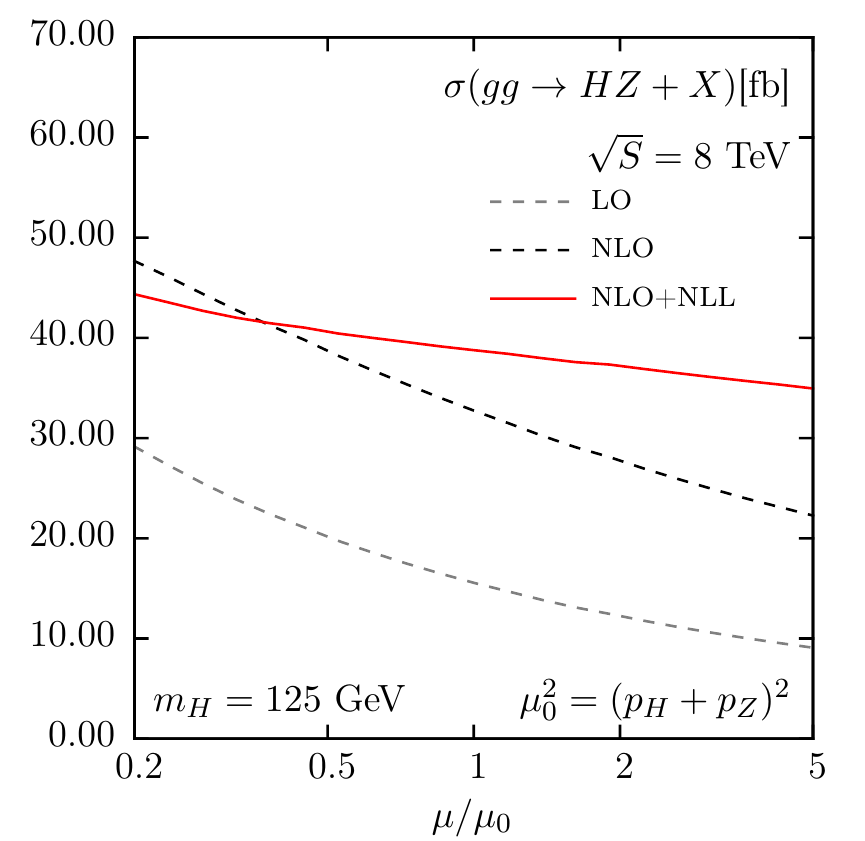} &
      \includegraphics[height=.35\textheight]{%
        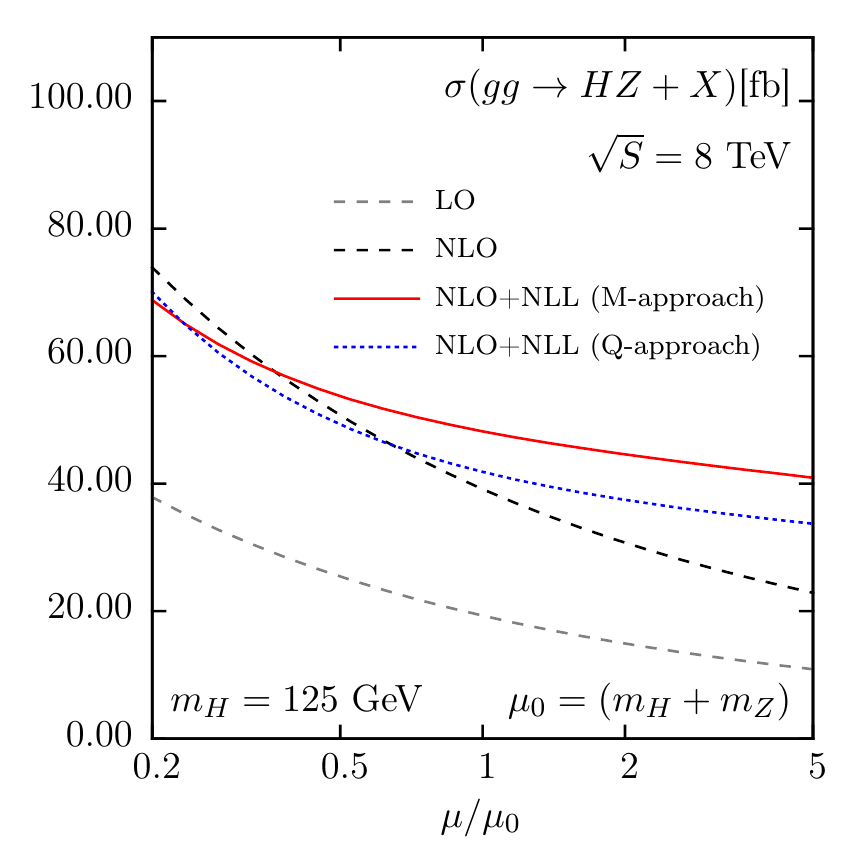}\\
      \hspace*{1.6em} (a) & \hspace*{1.9em} (b)
    \end{tabular}
    \parbox{.9\textwidth}{
      \caption[]{\label{fig:nll8}\sloppy Total inclusive cross section
        at $\sqrt s =8$\,TeV due to gluon-induced $\zh$ production
        at \lo{} (lower dashed; gray), \nlo{} (upper dashed; black), and
        \nlonll{} (solid; red). (a) \qapproach{} with central scale
        choice $\mu_0^2=Q^2$, (b) \mapproach{} with central scale choice
        $\mu_0^2=\qnt^2$. Also shown in (b) is the \nlonll{} result from
        the \qapproach{} for the central scale choice $\mu_0^2=\qnt^2$
        (dotted; blue).  }}
  \end{center}
\end{figure}
\begin{figure}
  \begin{center}
    \begin{tabular}{cc}
      \includegraphics[height=.35\textheight]{%
        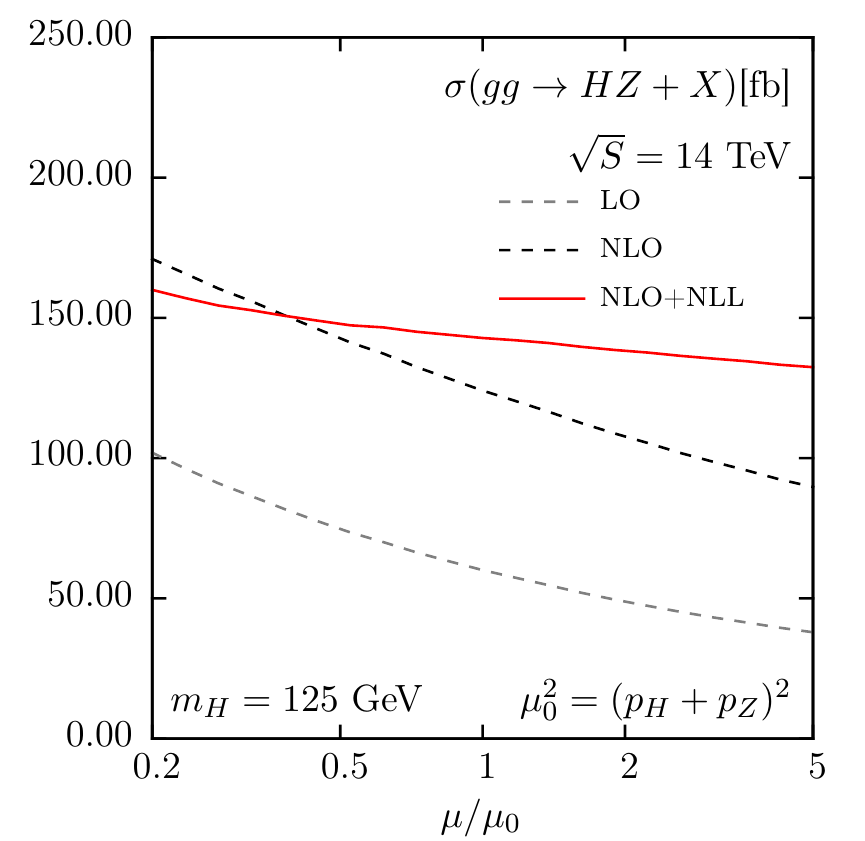} &
      \includegraphics[height=.35\textheight]{%
        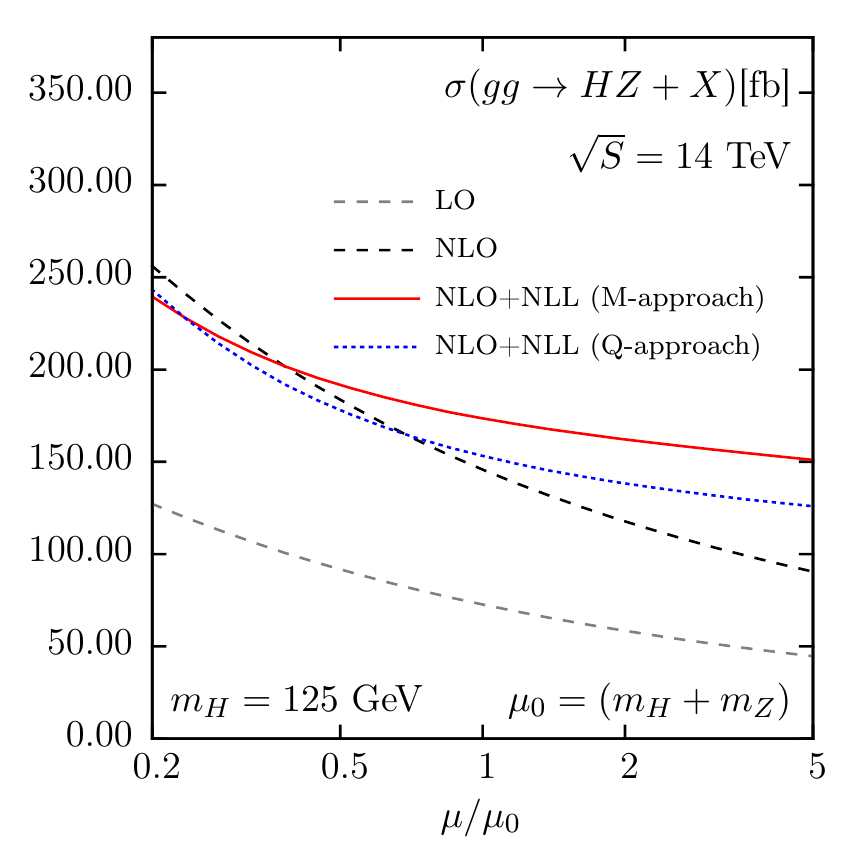}\\
      \hspace*{1.6em} (a) & \hspace*{1.9em} (b)
    \end{tabular}
    \parbox{.9\textwidth}{
      \caption[]{\label{fig:nll14}\sloppy Same as \fig{fig:nll8}, but
        for $\sqrt{s}=14$\,TeV. }}
  \end{center}
\end{figure}
%
%


\fig{fig:nll8} shows the \nlonll{} result for 8\,TeV collisions for the
gluon-initiated total inclusive cross section upon simultaneous
variation of the unphysical scales, $\mu\equiv \muF=\muR$, together with
the fixed order prediction at \lo{} and \nlo{}, for (a) the \qapproach{}
and (b) the \mapproach. As explained above, all curves have been
evaluated with \nnlo{} \pdf{}s. As expected, resummation significantly
decreases the scale dependence in both approaches. In accordance with
the supposition that the ``natural'' scale for the partonic process is
$Q^2=(p_H+p_Z)^2$, the absolute as well as the relative scale
uncertainty is considerably lower for the \qapproach{}. This is mostly
due to the fact that the fixed-order results exhibit a much stronger
scale dependence upon variation of $\mu$ relative to the fixed scale
$\qnt$. For example, while the \nlo{} (\lo) cross section varies by
$\pm$39\% ($\pm$65\%) when $\mu$ is changed by a factor of five around
the $\zh$ invariant mass, this variation is $\pm$66\% ($\pm$70\%) when
the scale is varied by the same factor around the absolute threshold
$\qnt=\mhiggs+\mz$. The resummed cross section, on the other hand,
varies only by $\pm$12\% in the \qapproach{} with $\mu_0^2=Q^2$, and by
$\pm$29\% in the \mapproach{} with $\mu_0=\qnt$. For comparison,
\fig{fig:nll8} also shows the \nlonll{} result obtained through the
\qapproach, but adopting $\mu_0=\qnt$ as the central resummation
scale. The scale variation is larger ($\pm$43\%) than in the
\mapproach{} with this scale choice. This is not unexpected, since when
$\mu_0=\qnt$ is chosen, the decisive logarithms of the ratio $W^2/\mu^2$
(cf. Eq.~\ref{eq:sudakov}) can get significantly larger for
$\genq^2=Q^2$ than for $\genq^2=\qnt^2$. Correspondingly, as the
absolute threshold region gets emphasized relative to configurations
with larger invariant mass by the choice $\mu_0=\mhiggs+\mz$, the effect
of resummation in the \mapproach{} is more enhanced than in the
\qapproach.

As shown in \fig{fig:nll14}, apart from an overall increase of the cross
section, the behavior of the curves for $pp$ collisions at
$\sqrt{s}=14$\,TeV is almost identical to the case of $8$\,TeV described
above.

Due to these considerations, and in view of the fact that the more
suitable scale for the process is considered to be the invariant mass of
the $\zh$ system, it is justified to base our final numerical
predictions on the results obtained through the \qapproach. The
comparison with the \mapproach{} results is reassuring since the two
results are quite compatible within the theoretical uncertainty, even
though the central value for the \mapproach{} is about 15\% larger than
for the \qapproach{} at $\mu_0=\mhiggs+M_Z$ and $\sqrt{s}=8$\,TeV.

Note also that for the central scale choice, we find an increase of the
total cross section at $\sqrt s =8$ TeV of about 18\% relative to the
\nlo{} result for the \qapproach{} (23\% for the \mapproach), which
might be a welcome feature for studies along the lines of
\citeres{Harlander:2013mla,Englert:2013vua}. For $\sqrt s=14$\,TeV the
corresponding increase of the cross section amounts to 15\%.

Before we present our updated prediction for the total inclusive $\zh$
cross section, let us discuss a further addition to the $\gghz$
subprocess which will, not least, allow us to validitate our estimate of
the theoretical uncertainty due to missing higher order terms, which
will be adopted in Section\,\ref{sec:update}. Since the final state in
the $\gghz$ process is a colour singlet, the radiative factors resumming
large logarithmic contributions are process independent and consist only
of $\Delta_g$ factors, cf.\ \eqn{eq:resumm}. Therefore we can easily
extend our analysis to include the \nnll{} terms in the radiative factor
$\Delta_g$, see \eqn{eq:sudakov}. Note that we will neglect the effect
of the \nnlo{}-term of the hard coefficient ${\cal C}_W$. The
calculation of this term will be only possible when the \nnlo{} result
for the $\gghz$ cross section is known.\footnote{Note, however, that the
  hard coefficient at \nnlo{} is given by the corresponding
  coefficient for the production of a pseudo-scalar
  particle\cite{deFlorian:2007sr}, plus higher order corrections to the
  reducible terms $\propto \sigma_\text{(virt,red)}$,
  cf.\,footnote\,\ref{fn:c1gga}. } Therefore we do not consider our
predictions including the \nnll{} radiative factor as final. Instead we
use them to validate the stability of the \nlonll{} result.

Our \nnll{} resummed results are matched to the \nlo{} cross
sections. Alternatively, we could have matched the \nnll{} result to an
approximation of the \nnlo{} cross section, consisting of the \nlo{}
result and soft terms of \nnlo{} order. However, in this way no
essential new information is generated since the soft terms are obtained
from an expansion of the resummed result up to \nnlo{}. The difference
between the \nnll{} result matched to \nlo{} and the same result matched
to the approximated \nnlo{} is of ${\cal O}(1/N)$ and as such beyond the
accuracy of resummation methods which do not control these
power-suppressed terms.


%
\begin{figure}
  \begin{center}
    \begin{tabular}{cc}
      \includegraphics[width=.45\textwidth]{%
        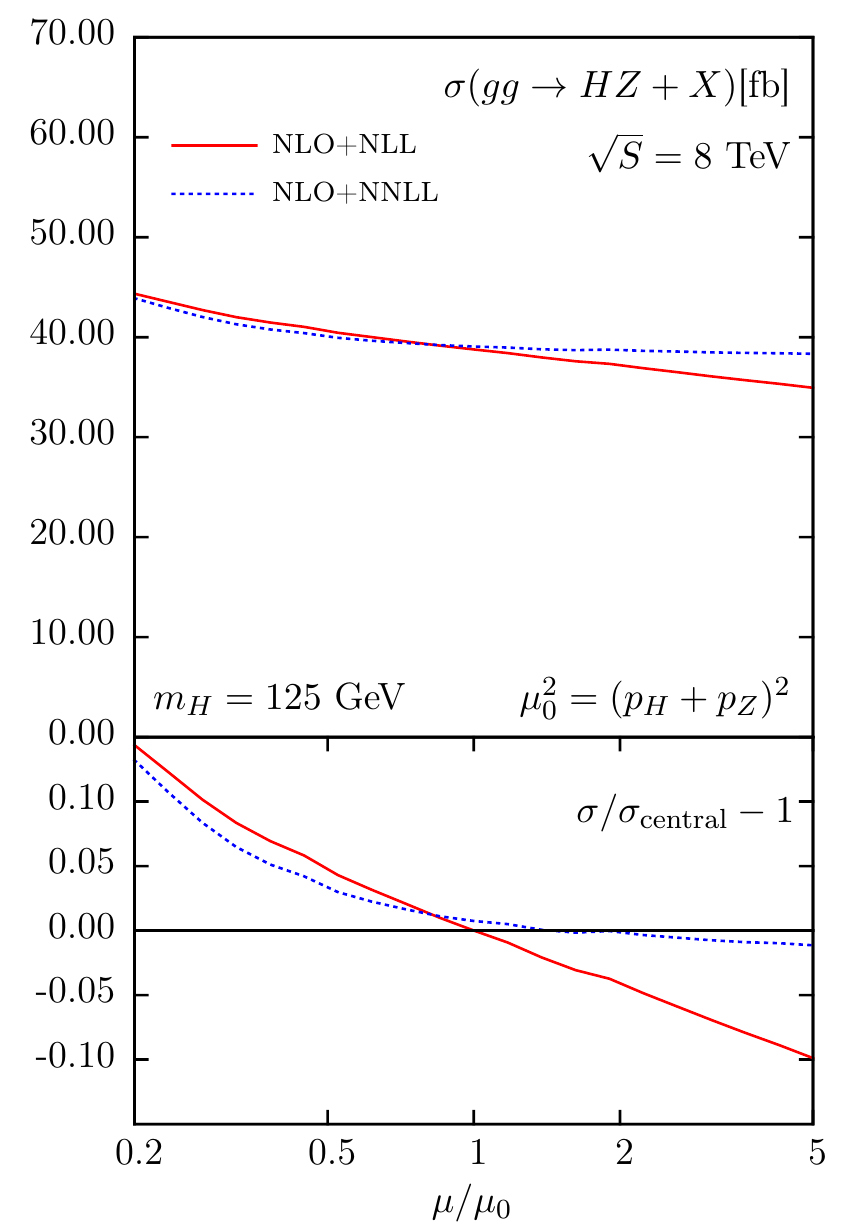}
      &
      \includegraphics[width=.45\textwidth]{%
        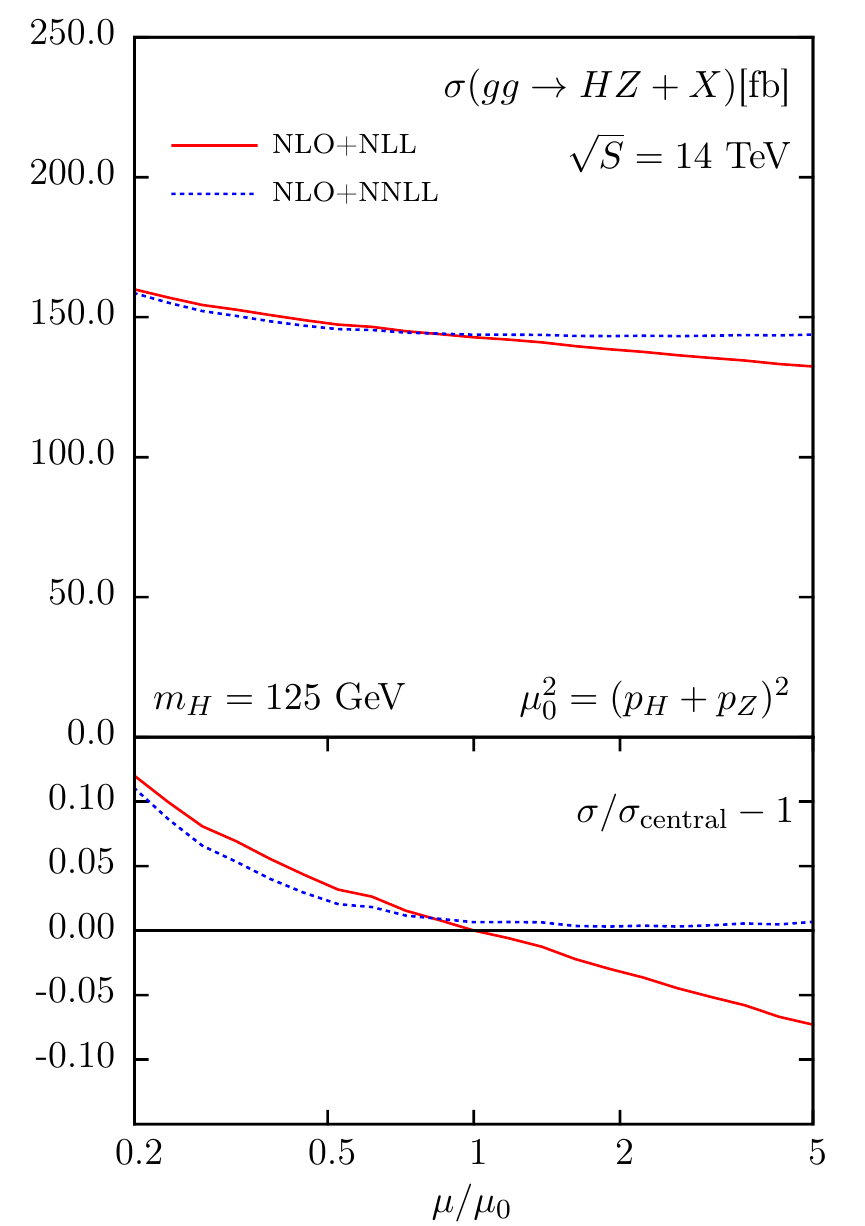}\\ 
      \hspace*{1.5em} (a) & \hspace*{1.5em} (b)
    \end{tabular}
    \parbox{.9\textwidth}{
      \caption[]{\label{fig:nnll8} Total inclusive cross section at
        (a)\,$\sqrt s =8$\,TeV and (b)\,$\sqrt{s}=14$\,TeV due to
        gluon-induced $\zh$ production at \nlonll{} (solid; red) and
        \nlo+\nnll{} (dotted; blue) in the \qapproach{}.  Upper plot:
        absolute values; lower plot: relative deviation to
        $\sigma_\text{central}\equiv \sigma_{\gghz}(\mu^2=Q^2)$, where
        $Q$ is the invariant mass of the $\zh$ system, and
        $\mu=\muF=\muR$ is the renormalization/factorization scale.}}
  \end{center}
\end{figure}
%


\fig{fig:nnll8} compares the \nlonll{} to the \nlo+\nnll{} result in the
\qapproach{}.  While the upper plot shows the absolute values, in the
lower plot the curves encode the relative deviation from what we will
use as our central prediction for $\sigma_{\gghz}$ in the next section,
namely the \nlonll{} cross section evaluated at the central value of the
renormalization and the factorization scale, $\mu^2\equiv
\muR^2=\muF^2=Q^2\equiv (p_H+p_Z)^2$.  We observe that the scale
dependence of the \nlo+\nnll{} is slightly smaller than for \nlonll{},
as it is expected due to the higher perturbative orders that are taken
into account. However, the \nlo+\nnll{} should only be considered as an
indicator of the size of possible higher order effects, since sub-leading
terms in $1/N$ as well as hard contributions at order $\alpha_s^3$ may
still give sizable contributions.  Therefore, we base our theoretical
prediction for the total cross section purely on the \nlonll{} result
which leads to a slightly more conservative error
estimate.\footnote{Note that any reasonable error band derived from the
  \nlonll{} result by varying the scale $\mu$ (e.g. within the interval
  $[1/3,3]\mu_0$) would cover the analogous band derived from the
  \nlo+\nnll{} result. This behavior provides quite some confidence in
  the perturbative stability of the \nlonll{} result.}


\section{The total inclusive cross section for $\zh$ production}
\label{sec:update}

In this section, we are going to present updated numbers for the total
inclusive cross section of associated $\zh$ production at the \lhc{}
which, in addition to all previously known effects, takes into account
the results of this paper for the resummed $\gghz$ contributions.  We
derive the updated prediction by evaluating the non-$\gghz$
contributions of $\zh$ production and their scale uncertainties using
{\tt vh@nnlo}\cite{Brein:2012ne,Harlander:2013mla}, and adding our new
results for the $\gghz$ component.

The latter are evaluated along the lines of \citere{Heinemeyer:2013tqa},
where for the central value, the renormalization and factorization scale
is set to the invariant mass of the $\zh$ system, $\muR=\muF=Q$, while
the theoretical error is determined by varying these two scales
simultaneously within the interval $[1/3,3]Q$.  The results for four
different energies and three different values of the Higgs mass are
shown in Table\,\ref{tab:gghz}. For comparison, we have also included
the numbers of the \nlo{} prediction.  The most notable effect of the
\nlonll{} result for $\gghz$ is the reduction of the theoretical error
due to scale variation by a factor of three to four. Also the central
value of the total inclusive rate increases by about 18\% for
$\sqrt{s}=8$\,TeV, and by about 15\% at 13 and 14\,TeV.

Let us now see how these effects translate to the total inclusive cross
section for the process $pp\to\zh$. In order to obtain the central value
of the total cross section, we add the central values of the $\gghz$ and
non-$\gghz$ components. The scale error of the latter is evaluated along
the lines of \citere{Heinemeyer:2013tqa} by varying $\muF$ and $\muR$
{\it independently} within the interval $[1/3,3]Q$, and added linearly
to the scale errors of the $\gghz$ contribution.  Our final prediction
is displayed in Table\,\ref{tab:xsec} for the same energies and Higgs
masses as before. For comparison, we have again included the numbers
based on the \nlo{} prediction of the $\gghz{}$ process which correspond
to the current recommended numbers of the {\abbrev
  LHC-HXSWG}\cite{Heinemeyer:2013tqa,Dittmaier:2012vm,Dittmaier:2011ti,
  hxswg}.\footnote{The current numbers of the {\abbrev LHC-HXSWG} at
  $\sqrt{s}=8$\,TeV have a symmetrized error, thus they slightly differ
  from the ones listed in Table\,\ref{tab:xsec}.} The effect of the
reduction of the scale variation in the $\gghz$ subprocess on the total
cross section is still significant, amounting to more than a factor of
two in many cases. This reflects the dominance of the $\gghz$ error in
the total uncertainty of this process, which can be inferred also by
comparing the scale error of the $\zh$ cross section to the one for
$W\!H$ production. The central value of the total inclusive rate
increases by about 1.5\% for $\sqrt{s}=8$\,TeV, and by about 2\% at 13
and 14\,TeV.


\renewcommand{\arraystretch}{1.3}
\begin{table}
\begin{center}
\begin{tabular}{|r|c||c|c|c|c|}
 \hline \multicolumn{1}{|c|}{$\sqrt{s}$} & $M_H$ & 
\multicolumn{4}{c|}{$\sigma_{gg\to H\!Z}$} \\
\hline
\multicolumn{1}{|c|}{[TeV]} & [GeV] & 
\multicolumn{2}{c|}{\nlo} &
\multicolumn{2}{c|}{\nlonll} \\
\cline{3-6}
& & [pb] & [\%] & [pb] & [\%] \\
\hline
8.00 & 125.0 & $ 0.0328 $&$ ^{+ 30} _{-
   23}$ & $ 0.0389$&$ ^{+ 8.1} _{- 6.5}$ \\ 8.00 & 125.5 & $ 0.0325 $&$
 ^{+ 30} _{- 23}$ & $ 0.0386$&$ ^{+ 8.1} _{- 6.5}$ \\ 8.00 & 126.0 & $
 0.0324 $&$ ^{+ 30} _{- 23}$ & $ 0.0384$&$ ^{+ 8.2} _{- 6.6}$ \\ 13.0
 & 125.0 & $ 0.1057 $&$ ^{+ 26} _{- 21}$ & $ 0.1220$&$ ^{+ 7.0} _{-
   5.2}$ \\ 13.0 & 125.5 & $ 0.1049 $&$ ^{+ 26} _{- 20}$ & $ 0.1211$&$
 ^{+ 7.3} _{- 5.2}$ \\ 13.0 & 126.0 & $ 0.1044 $&$ ^{+ 26} _{- 21}$ &
 $ 0.1205$&$ ^{+ 6.7} _{- 5.5}$ \\ 13.5 & 125.0 & $ 0.1149 $&$ ^{+ 26}
 _{- 20}$ & $ 0.1325$&$ ^{+ 7.1} _{- 5.2}$ \\ 13.5 & 125.5 & $ 0.1141
 $&$ ^{+ 26} _{- 20}$ & $ 0.1316$&$ ^{+ 7.2} _{- 5.0}$ \\ 13.5 &
 126.0 & $ 0.1135 $&$ ^{+ 26} _{- 20}$ & $ 0.1308$&$ ^{+ 7.1} _{-
   5.0}$ \\ 14.0 & 125.0 & $ 0.1243 $&$ ^{+ 26} _{- 20}$ & $ 0.1431$&$
 ^{+ 7.1} _{- 4.9}$ \\ 14.0 & 125.5 & $ 0.1237 $&$ ^{+ 26} _{- 20}$ &
 $ 0.1424$&$ ^{+ 7.3} _{- 5.0}$ \\ 14.0 & 126.0 & $ 0.1228 $&$ ^{+ 26}
 _{- 20}$ & $ 0.1414$&$ ^{+ 7.3} _{- 4.9}$
\\ \hline
\end{tabular}
\parbox{.9\textwidth}{%
\caption[]{\label{tab:gghz} The $\gghz$ component to the total inclusive
  $\zh$ production cross section for various center-of-mass energies
  $\sqrt{s}$ and Higgs masses $\mhiggs$. The third/fourth column
  displays the \nlo{} result, while the fifth/sixth column shows the
  \nlonll{} prediction (based on the \qapproach) calculated in this
  paper. The uncertainties (column 4 and 6) are due to scale
  variation. }}
\end{center}
\end{table}


\renewcommand{\arraystretch}{1.3}
\begin{table}
\begin{center}
\begin{tabular}{|r|c||c|c|c|c|}
 \hline \multicolumn{1}{|c|}{$\sqrt{s}$} & $M_H$ & 
\multicolumn{4}{c|}{$\sigma_{pp\to H\!Z}$} \\
\hline
\multicolumn{1}{|c|}{[TeV]} & [GeV] & 
\multicolumn{2}{c|}{{\footnotesize incl.}} &
\multicolumn{2}{c|}{{\footnotesize incl.}} \\[-.6em]
\multicolumn{1}{|c|}{} &  & 
\multicolumn{2}{c|}{{\footnotesize $gg$ @ \nlo}} &
\multicolumn{2}{c|}{{\footnotesize $gg$ @ \nlonll}} \\
\cline{3-6}
& & [pb] & [\%] & [pb] & [\%] \\
\hline 
8.00 & 125.0 & $ 0.4157 $&$ ^{+ 3.1} _{-
   2.8}$ & $ 0.4217$&$ ^{+ 1.5} _{- 1.5}$ \\ 8.00 & 125.5 & $ 0.4104 $&$
 ^{+ 3.1} _{- 2.8}$ & $ 0.4165$&$ ^{+ 1.5} _{- 1.5}$ \\ 8.00 & 126.0 & $
 0.4054 $&$ ^{+ 3.2} _{- 2.8}$ & $ 0.4114$&$ ^{+ 1.5} _{- 1.6}$ \\ 13.0 &
 125.0 & $ 0.8696 $&$ ^{+ 3.8} _{- 3.8}$ & $ 0.8859$&$ ^{+ 1.6} _{-
   2.0}$ \\ 13.0 & 125.5 & $ 0.8594 $&$ ^{+ 3.8} _{- 3.8}$ & $ 0.8757$&$
 ^{+ 1.7} _{- 1.9}$ \\ 13.0 & 126.0 & $ 0.8501 $&$ ^{+ 3.8} _{- 3.9}$ & $
 0.8663$&$ ^{+ 1.6} _{- 2.1}$ \\ 13.5 & 125.0 & $ 0.9190 $&$ ^{+ 3.9}
 _{- 3.8}$ & $ 0.9366$&$ ^{+ 1.6} _{- 2.0}$ \\ 13.5 & 125.5 & $ 0.9085
 $&$ ^{+ 3.8} _{- 3.8}$ & $ 0.9259$&$ ^{+ 1.6} _{- 2.0}$ \\ 13.5 & 126.0
 & $ 0.8988 $&$ ^{+ 3.8} _{- 3.9}$ & $ 0.9162$&$ ^{+ 1.5} _{- 2.0}$
 \\ 14.0 & 125.0 & $ 0.9690 $&$ ^{+ 4.0} _{- 3.9}$ & $ 0.9878$&$ ^{+ 1.7}
 _{- 2.0}$ \\ 14.0 & 125.5 & $ 0.9574 $&$ ^{+ 4.0} _{- 3.9}$ & $ 0.9761$&$
 ^{+ 1.7} _{- 2.0}$ \\ 14.0 & 126.0 & $ 0.9465 $&$ ^{+ 4.1} _{- 3.9}$ & $
 0.9652$&$ ^{+ 1.7} _{- 2.0}$
\\ \hline
\end{tabular}
\parbox{.9\textwidth}{%
\caption[]{\label{tab:xsec}Total inclusive cross section for the process
  $pp\to \zh$ for various center-of-mass energies $\sqrt{s}$ and Higgs
  masses $\mhiggs$. The third/fourth column includes the $\gghz$
  subprocess through \nlo{}, while the fifth/sixth column includes the
  \nlonll{} prediction (based on the \qapproach) for the $\gghz$
  subprocesses as calculated in this paper. The uncertainties (column 4
  and 6) are due to scale variation. \pdf+$\alpha_s$ uncertainties are
 at the few-percent level and are not listed here; they can be taken
 over from the {\abbrev LHC-HXSWG} recommendations \cite{hxswg}.}}
\end{center}
\end{table}



\section{Conclusions}\label{sec:conclusions}

We have presented the threshold resummed cross section for gluon-induced
$\zh$ production through \nll{}, matched to the \nlo{} result. Two
approaches were pursued: in the \qapproach, logarithms relative to the
threshold of producing a $\zh$ pair of invariant mass $Q$ were resummed,
while in the \mapproach, the resummation was performed relative to the
absolute threshold of the $\zh$ system, $M=M_H+M_Z$. It was argued that,
due to the possibility of a more appropriate choice for the
renormalization and the factorization scale, the \qapproach{} leads to
perturbatively more stable and thus more precise predictions for the
$\gghz$ cross section.  In order to underline the reliability of our
prediction, we considered the universal terms arising at \nnll{} and
found complete compatibility with the \nlonll{} result.

The \nlonll{} result was found to exhibit a strongly reduced scale
variation relative to the \nlo{} result, while the central value was
increased. This motivates a re-evaluation of the total inclusive $\zh$
cross section, including the previously known \nnlo{} corrections to the
``Drell-Yan-like'' contributions\cite{Brein:2003wg} as well as top-loop
induced effects at $\alpha_s^2$\cite{Brein:2011vx}. The central value of
the new prediction for $\sigma(pp\to HZ)$ is slightly larger than the
value currently suggested by the {\abbrev LHC-HXSWG}, and the
theoretical error due to scale variation is reduced by roughly a factor
of two.


\paragraph{Acknowledgments.}
We thank Stefan Liebler for assistance with the $\gghz$ subroutines of
{\tt vh@nnlo}, and Fabio Maltoni for useful discussions.  This research was
supported by the Munich Institute for Astro- and Particle Physics
(MIAPP) of the DFG cluster of excellence ``Origin and Structure of the
Universe'', and by the DFG contract HA\,2990/5-1. The support by Polish
National Science Centre grant, project number DEC-2011/01/B/ST2/03643,
is gratefully acknowledged.


\end{document}